\documentclass[reprint,aps,groupedaddress,showpacs]{revtex4-1}
\usepackage{graphicx}
\usepackage{amsmath,amssymb,lmodern}
\usepackage{amsfonts}
\usepackage{amssymb}
\usepackage{bm}
\begin{document}
%\begin{CJK}{GBK}{song}
%\title{Spin Chern Pumping Effect in Topological Insulators}
\title{Spin Chern Pumping from the Bulk of Two-Dimensional Topological Insulators}

\author{M. N. Chen$^1$}
\author{L. Sheng$^{1}$}
\email{shengli@nju.edu.cn}
\author{R. Shen$^{1}$}
\author{D. N. Sheng$^2$}
\author{D. Y. Xing$^{1}$}
\email{dyxing@nju.edu.cn}
\affiliation{$^1$National Laboratory of Solid State Microstructures, Department of Physics, and  Collaborative Innovation Center of Advanced Microstructures, Nanjing University, Nanjing 210093, China\\
$^2$ Department of Physics and Astronomy, California State
University, Northridge, California 91330, USA}
%$^3$ Collaborative Innovation Center of Advanced Microstructures, Nanjing 210093, China}

\begin{abstract}
Topological insulators (TIs) are a new quantum state of matter discovered
recently, which are characterized by unconventional bulk
topological invariants. Proposals for practical applications
of the TIs are mostly based upon their metallic surface or edge states.
%: either the $Z_2$ index or spin Chern numbers.
%Unlike the first Chern number underlying the quantum Hall effect,
%up to now these topological
%invariants have not been directly measurable.
Here, we report the theoretical discovery of
a bulk quantum pumping effect
in a two-dimensional TI electrically modulated in
adiabatic cycles. In each cycle, an amount of spin
proportional to the sample width can be
pumped into a nonmagnetic electrode, which is attributed to
nonzero spin Chern numbers $C_{\pm}$.
Moreover, by using a half-metallic electrode, universal quantized
charge pumping conductivities $-C_{\pm}e^2/h$ can be measured.
This discovery paves the way for direct investigation
of the robust topological properties of the TIs.
\end{abstract}

 \maketitle

\section{INTRODUCTION}

Topological transport phenomena have been attracting a great
deal of interest, because they exhibit universal properties that
are insensitive to perturbations and independent of
material details. A classical example
of such a transport phenomenon is the integer quantum Hall (IQH)
effect in two-dimensional (2D) electron systems,
first discovered in 1980,~\cite{Klitzing} which is characterized
by an integer quantization of
the Hall conductivity in unit of $e^2/h$.
The IQH effect has been
observed in a large variety of materials, ranging from
traditional semiconductors, to oxides,~\cite{IQHE_Oxides}
graphene,~\cite{IQHE_graphene} and topological insulators (TIs).~\cite{IQHE_topo}
Laughlin~\cite{Laughlin} interpreted the IQH effect
in terms of an adiabatic charge pump.
Thouless, Kohmoto, Nightingale, and Nijs~\cite{Thouless0} established a
relation between the quantized Hall conductivity of the IQH system and
a topological invariant, the first Chern number.
Thouless and Niu~\cite{Thouless1,Thouless2}
also related the amount of charge
pumped in a 1D charge pump to the Chern number.

A variant of the IQH effect, the quantum spin Hall  (QSH) effect,
was proposed recently,~\cite{QSH1,QSH2} which has been experimentally
realized in HgTe quantum wells~\cite{HgTe} and InAs/GaSb
bilayers.~\cite{RRDu} Extension of the idea of the QSH effect has led to
the discovery of 3D TIs.~\cite{TI1,TI2,TI3,TI4}
A QSH system, which is also called a 2D TI,
has an insulating band gap in the bulk and a pair of gapless
helical edge states at the sample boundary.
When the electron spin is conserved, a QSH system can be
viewed as two independent IQH systems without
Landau levels.~\cite{Haldane} Different from the charge, the
spin does not obey a fundamental conservation law.
In general, when the spin conservation is absent,
unconventional topological invariants, either
the $Z_2$ index~\cite{Z2index} or
the spin Chern numbers,~\cite{spinch1,spinch2,spinch4}
are needed to describe the QSH systems.
The time-reversal (TR) symmetry is considered to be a
prerequisite for the QSH effect, which protects
both the $Z_2$ index and gapless nature of the edge states.
However, based upon the spin Chern numbers, it was
shown that the bulk topological properties remain
intact even when the TR symmetry is broken.~\cite{spinch4}
This finding evokes interest to pursue direct investigation
and possibly utilization
of the robust topological properties of the TIs,
besides using their symmetry-protected gapless edge states
which are more fragile in realistic environments.

Unlike the first Chern number underlying the IQH
systems, which is embedded into the Hall conductivity, up to now the
topological invariants in the TIs have not been  directly
observable. Several experimental methods were proposed, but have not
been realized. One  was to measure
the topological magnetoelectric effect,~\cite{Magneto1,Magneto2}
for which experimental complexities exist.~\cite{Magneto2}
Fu and Kane~\cite{Z2pump} put forward
an abstract 1D model, in which the spin pumping
was related to the $Z_{2}$ index in the limit of weak
coupling. However,
how this fictitious model could be
implemented is still unknown. Furthermore, from the
viewpoint of application, generalization of the idea of the $Z_{2}$ pump
to higher dimension is meaningless, because according to the $Z_{2}$
theory,~\cite{Z2pump} only the states at the
TR-invariant point of the Brillouin zone can
contribute to the spin pumping, and so the pumping rate
cannot be enhanced by increment of dimension.
In a recent work,~\cite{SCPump} the more general case of finite coupling between
the pump and electrode is investigated by using the scattering matrix method.
It was found that the spin pumping in the model of Fu and Kane
can survive finite scattering of
magnetic impurities, and so may be attributed to the spin Chern numbers
rather than the $Z_2$ index.
Some other authors~\cite{RealizeZ2_0, RealizeZ2_1}
proposed to pump quantized charge
through the helical edge states by precessing a magnet
covering the edge of a 2D TI, so that
the number of gapless edge channels can be counted through electrical
measurement.
This method is indirect, in the sense that
the topological invariants are intrinsic properties of the bulk
electron wavefunctions, which do
not immediately determine the charge pumping in the
edge channels.
%Interestingly, it has been shown that
%the spin pumping effect in the model reveals the bulk
%topological invariants of a corresponding
%2D TI.~\cite{Z2pump,SpinChernPump}

Here we  predict an intriguing bulk
topological pumping effect, directly
driven by nonzero spin Chern numbers,
in a QSH system electrically
modulated in adiabatic cycles.
As a consequence of the topological
spectral flows of the spin-polarized Wannier
functions (SPWFs) in the bulk of the system,
spin can be pumped into a nonmagnetic
electrode continuously without net charge transfer.
The total amount of spin pumped per cycle is
proportional to the (cross-section) width
of the sample, and
insensitive to the material parameters
and spin-mixing effect due to
the Rashba spin-orbit coupling.
This electrical spin pump establishes a basis,
on which spintronic applications
taking advantage of
the robust topological properties of the TIs
can be developed.
Especially, if a half-metallic electrode
with spin polarization parallel
(or antiparallel) to the $z$-axis is used, a quantized charge
pumping conductivity, $-C_{+}e^2/h$ (or $-C_{-}e^2/h$),
can be measured by electrical means,
demonstrating a way to observe
the spin Chern numbers $C_{\pm}$ directly.

%By viewing
%a momentum component, $k_y$, as a parameter, the topological properties
%of the system are described by the spin Chern numbers $C_{\pm}(k_y)$
%defined for two variables, $k_x$ and time $t$. We find that
%$C_{\pm}(k_y)=\pm 1$ in the region of $-k_{y}^c<k_y<k_{y}^c$,
%and vanish elsewhere. Correspondingly, when an metallic lead is connected
%to the system along the $x$ direction, a quantized spin $\hbar$
%is pumped to the lead per cycle at each $k_y$ between
%$\pm k_{y}^c$, and the total spin pumped is
%proportional to the sample width. It is also shown that turning on
%the Rashba spin-orbit coupling leads to a small deviation from the
%quantized value.
\section{Spin Chern numbers and SPWFs}
Let us consider a 2D model Hamiltonian $H_{P}=H_{0}+H_{1}$
with
\begin{equation}
H_{0}=v_{\mbox{\tiny F}}\left[k_x\hat{s}_z\hat{\sigma}_x-\left(k_{y}+eA(t)\right)
\hat{\sigma}_y\right]-M(t)\hat{\sigma}_{z}\ .
\label{Hamil}
\end{equation}
Here $(-e)$ is the electron charge, $\bf k$ is the 2D momentum,
$A(t)=A_{0}\sin(\omega_{0}t)$ is the vector potential of an
$ac$ electric field $-E_{0}\cos(\omega_{0}t)$
applied along the $y$ direction with $A_{0}=E_{0}/\omega_{0}$
and frequency $\omega_{0}>0$ being designated,
and $M(t)=M_{0}\cos(\omega_{0}t)$.  This model can describe both the QSH materials,
the HgTe quantum wells,~\cite{BHZModel,BHZModel1,DualGate1} and InAs/GaSb
bilayers,~\cite{DualGate2}  in the linear order in momentum.
For definity of discussion, we confine ourselves to
the HgTe quantum wells, for which $\hat{s}_{\alpha}$
with $\alpha=x,y,z$ are the Pauli matrices for
spin, and $\hat{\sigma}_{\alpha}$ for the electron and hole bands.
As will be discussed below,
the time-dependent mass term $M(t)$ can be
induced by varying the voltages of the dual gates.
$H_{1}$ represents the Rashba
spin-orbit coupling~\cite{HgTeRashba}
\begin{equation}
H_{1}=\frac{R_{0}}{2}(\hat{1}+\hat{\sigma}_{z})
[\hat{s}_{y}k_{x}-\hat{s}_{x}(k_{y}+eA(t))]\ .
\label{Rashba}
\end{equation}
To the linear order in momentum, the Rashba spin-orbit coupling
is nonvanishing only in the electron band.~\cite{HgTeRashba}

Within the adiabatic approximation,
%the eigenenergy
%of $H_{P}$ for $R_{0}=0$ can be obtained as
%\begin{equation}
%E_{\pm}^{v(c)}=+(-)\sqrt{v^2_{\mbox{\tiny F}}\tilde{k}^2+M^2(t)}\ ,
%\end{equation}
%where $\tilde{k}^2=k_{x}^2+\tilde{k}_{y}^2$ with
%$\tilde{k}_{y}=(k_y+eA(t))$.
for a bulk sample there exists a finite energy gap
between the conduction and valence bands for $\omega_{0}t\neq\pi/2$
or $3\pi/2$. At $\omega_{0}t=\pi/2$ and $3\pi/2$, the conduction and valence
bands touch at $k_{x}=0$ and $k_{y}=k_{y}^{c}$ or $-k_{y}^{c}$
with $k_{y}^{c}=e\vert A_{0}\vert=e\vert E_{0}\vert/\omega_{0}$.
To clarify the topological
properties underlying the spin/charge pumping,
we consider $k_{y}$ as
a parameter, and calculate the spin Chern numbers $C_{\pm}$
in the standard way,~\cite{spinch4}
on the torus of the two variables
$k_{x}\in (-\infty,\infty)$ and $t\in[0, T)$ with $T=2\pi/\omega_{0}$
as the period. The
spin Chern numbers are obtained as
\begin{equation}
C_{\pm}=\pm\mbox{sgn}(E_{0}M_{0})\ ,
\label{sChern}
\end{equation}
for $\vert k_y\vert<k_{y}^c$, and vanish elsewhere.
Not surprisingly, the band touching points $k_{y}=\pm k_{y}^{c}$
serve as the critical points.

\begin{figure}
\includegraphics[width=3.0in]{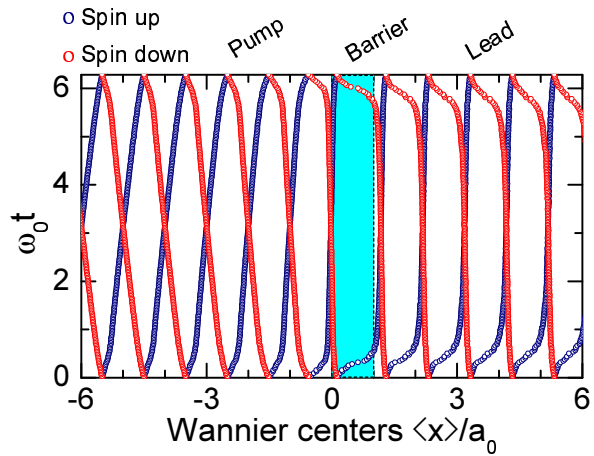}
\caption{Plot of the centers of mass of the SPWFs (horizontal axis)
as functions of $\omega_{0}t$ (vertical
axis). The parameters are taken to be
$k_{y}=0.4k_{y}^{c}$, $M_{0}=v_{\mbox{\tiny F}}eA_{0}=
\hbar R_{0}/a_{0}=0.1t_{0}$, $V_{0}=0.3t_{0}$, and $d=a_{0}$,
with $a_{0}$ as the lattice constant and $t_{0}=\hbar v_{\mbox{\tiny F}}/a_{0}$
($v'_{\mbox{\tiny F}}=v_{\mbox{\tiny F}}$)
as the hopping integral of the tight-binding Hamiltonian.
} \label{Fig_1}
 \end{figure}
We now consider a system consisting of
a pump for $x < 0$,
an electrode for $x > d$, and a potential barrier
in between. The total Hamiltonian of
the system reads
\begin{equation}
H=\left\{
\begin{array}{ll}
H_{P}&\mbox{ }(x<0)\\
H_{E}+V_{0}\hat{\sigma}_{z}&\mbox{ }(0<x<d)\\
H_{E}&\mbox{ }(x>d)
\end{array}
\right.\ ,
\label{Htot}
\end{equation}
where $H_{P}$ has been given above, and
$H_{E}=v'_{\mbox{\tiny F}}k_x\hat{s}_z\hat{\sigma}_x$
is the Hamiltonian of the electrode. A possible experimental
setup for realizing this Hamiltonian
is explained in Appendix A in more details. In the barrier region,
the term $V_{0}\hat{\sigma}_{z}$ opens an insulating
gap of size $2V_{0}$, which accounts for
contact deficiencies between the pump and electrode.
The crucial role of the nonzero spin Chern numbers
in the spin/charge pumping process can be
visualized by using the SPWFs, which were first introduced
in Ref.\ \cite{SCPump}.
We construct a tight-binding Hamiltonian
for the effective
1D system at any given $k_y$
according to Eq.\ (\ref{Htot}), and diagonalize the total Hamiltonian
of the pump and electrode numerically.
Following the same procedure as calculating the spin Chern
numbers,~\cite{spinch4} the
space occupied by electrons is partitioned into two spin
sectors after diagonalizing the spin operator $\hat{s}_z$
in the occupied space.
%If the spin is conserved, the eigenvalues
%can have only two values:  $1$ or $-1$. When the spin conservation
%is broken weakly, a finite gap still exists in the eigen-spectrum
%of $Ps_{z}P$, which naturally divides the spectrum into two
%sectors:  spin-up and spin-down sectors.
%By definition, the corresponding
%eigenfunctions of $Ps_{z}P$ are the maximally spin-polarized states.
By definition, the states in the two spin sectors
are essentially the maximally spin-polarized states.
Then we construct the Wannier functions~\cite{Wannier,QiXL}
for the spin-up and spin-down sectors,
respectively, which are called the SPWFs.

The evolution of the centers of mass of the SPWFs for
$k_y=0.4k_{y}^{c}$ and $R_{0}=0.1v_{\mbox{\tiny F}}$
is shown in Fig.\ \ref{Fig_1}.
We see that the Wannier centers for the spin-up sector 
move right and those for the spin-down sector move left,
each center shifting on average a lattice constant per cycle.
Within the adiabatic approximation, 
time $t\in [0, T)$ plays the same role as the momentum
of an additional dimension,~\cite{Z2pump} namely, $k_{t}\in [0, T)$.
Therefore, when $k_y$ is considered as a parameter, 
the evolution of the Wannier functions of the effectively 1D system related to
various $k_x$ with time $t$ can be understood from the static properties of 
a 2D system associated with various $k_x$ and $k_t$.
In the general theory,~\cite{QiXL} the 
relationship between the Chern number and the spectral flows of the
Wannier functions in a 2D system has been established.
According to this theory, the average displacement  
of each of the centers of the SPWFs in the spin-up (spin-down) sector
with changing $k_t$ (or $t$) from $0$ to $T$, 
in units of the lattice constant, must equal to the
spin Chern number $C_{+}=1$ ($C_{-}=-1$). Therefore, the nontrivial transfer of the SPWFs
observed in Fig.\ 1 is a direct manifestation of the nonzero spin Chern numbers
$C_{\pm}=\pm 1$ in the pump (for $E_{0}M_{0}>0$).
More interestingly, we see that such spectral
flows can go across the finite barrier ($V_{0}d>0$),
and extend into the electrode,  even though the barrier
and electrode are
topologically trivial. Physically, because the system needs to 
recover its original eigenstates when each cycle ends, the 
nontrivial spectral flows
of the SPWFs in the TI need to constitute closed loops through
formation of edge states at the boundary,~\cite{HCLi2} or extend into
the electrode. However, localized edge states can not exist 
at the finite barrier due to quantum tunneling effect, 
so the transfer of the spectral flows of the SPWFs into 
the electrode occurs. This result will be
further confirmed by direct calculation based upon the scattering matrix
theory in the next section.

The SPWFs
are just another equivalent representation of the occupied space,
and so the counter spectral flows of the Wannier centers in the
two spin sectors represent the true movements of the electrons.
If the Rashba spin-orbit coupling were neglected, the Wannier
functions would be the eigenstates of $\hat{s}_{z}$.
The nontrivial spectral flows indicate that
at the given $k_y$, in each cycle a spin-up electron goes from the pump into
the electrode, and a spin-down electron moves oppositely.
Therefore, no net charge transfer occurs but a quantized
spin of $2(\hbar/2)$ is pumped into the electrode. When the small Rashba
spin-orbit coupling is turned on, while the topological
spectral flows remain intact, as seen from
Fig.\ \ref{Fig_1}, the spin polarizations of
the Wannier functions are
no longer fully parallel to the $z$-axis, and may also vary
with time. As a consequence, the amount of spin pumped per cycle
will deviate from the quantized value.\\
\\

\section{The Process of Spin Chern Pumping}
\subsection{Spin pumping for a nonmetallic electrode}
In general, the amount of the  spin pumped can be conveniently
calculated by using the scattering matrix formula.~\cite{SMat1,SMat2}
The $z$-component of the spin pumped
per cycle is given by~\cite{SMat1,SMat2}
\begin{widetext}
\begin{eqnarray}
\Delta s_{z}(k_y)=\frac{\hbar}{4\pi i}\oint_{T}
dt\Bigl(r^{*}_{\uparrow\uparrow}\frac{dr_{\uparrow\uparrow}}{dt}
-r^{*}_{\downarrow\downarrow}\frac{dr_{\downarrow\downarrow}}{dt}
-r^{*}_{\downarrow\uparrow}\frac{dr_{\downarrow\uparrow}}{dt}+
r^{*}_{\uparrow\downarrow}\frac{dr_{\uparrow\downarrow}}{dt}\Bigr)\ ,
\label{DS1}
\end{eqnarray}
where $r_{\alpha\beta}$ ($\alpha,\beta=\uparrow,\downarrow$)
is the reflection amplitude for an
electron at the Fermi energy incident from the spin-$\beta$
channel of the electrode and reflecting back into the spin-$\alpha$ channel.
In the following calculations,
the Fermi energy is set to be zero ($E_ F=0$), and
the Rashba spin-orbit coupling
is treated as a perturbation. As shown in Appendix B, to the linear order in $R_{0}$,
we obtain
\begin{equation}
r_{\uparrow\uparrow}=-\frac{\cos(2\theta)+i[\mbox{sh}(2\gamma_{0}d)-\sin(2\theta)
\mbox{ch}(2\gamma_{0}d)]}{\mbox{ch}(2\gamma_{0}d)-\sin(2\theta)
\mbox{sh}(2\gamma_{0}d)}+{\cal O}(\epsilon^2)\ ,
\label{Rupup0}
\end{equation}
\end{widetext}
\begin{equation}
r_{\downarrow\uparrow}=\frac{\epsilon}{2}\frac{\sin(2\theta)[1-\cos(2\theta)]}
{\mbox{ch}(2\gamma_{0}d)-\sin(2\theta)\mbox{sh}(2\gamma_{0}d)}
+{\cal O}(\epsilon^2)\ ,
\label{Rdnup0}
\end{equation}
and $r_{\downarrow\downarrow}=r_{\uparrow\uparrow}\vert_{2\theta\rightarrow(\pi-2\theta)}$ and
$r_{\uparrow\downarrow}=-r_{\downarrow\uparrow}\vert_{2\theta\rightarrow(\pi-2\theta)}$,
where $\gamma_{0}=V_{0}/\hbar v'_{\mbox{\tiny F}}$ and $2\theta=
\mbox{arg}[v_{\mbox{\tiny F}}(k_y+eA(t))+iM(t)]$. We note that
the dimensionless quantity $\epsilon=R_{0}/v_{\mbox{\tiny F}}$ appears as the
small expansion parameter. Since $r_{\downarrow\uparrow}$ and $r_{\uparrow\downarrow}$
are always real, the contributions from the third and fourth terms in Eq.\ (\ref{DS1})
vanish. Consequently,
\begin{equation}
\Delta s_{z}(k_y)=\frac{\hbar}{4\pi i}\oint_{T}\left(
r^{*}_{\uparrow\uparrow}dr_{\uparrow\uparrow}
-r^{*}_{\downarrow\downarrow}dr_{\downarrow\downarrow}\right)\ .
\label{DS2}
\end{equation}
This expression has a geometric
explanation: the amount of spin pumped per cycle equals to the difference
between the
areas enclosed by the directional
trajectories of $r_{\uparrow\uparrow}$ and
$r_{\downarrow\downarrow}$ on the complex plane,  multiplied by $\hbar/2\pi$.
Due to the relation $r_{\downarrow\downarrow}
=r_{\uparrow\uparrow}\vert_{2\theta\rightarrow(\pi-2\theta)}$,
yielding $r_{\downarrow\downarrow}=-\mbox{Re}(r_{\uparrow\uparrow})+i
\mbox{Im}(r_{\uparrow\uparrow})$,
the two terms in Eq.\ (\ref{DS2}) make an
equal contribution, so that we can focus on the first term.
While the expression (\ref{Rupup0}) for  $r_{\uparrow\uparrow}$
is independent of the Rashba spin-orbit coupling,
as will be shown soon, a combination of Eqs.\ (\ref{Rupup0})
and (\ref{Rdnup0}) allows us to evaluate
the amount of spin pumped
up to the second order in $R_{0}/v_{\mbox{\tiny F}}$.

\begin{figure}
\includegraphics[width=2.4in]{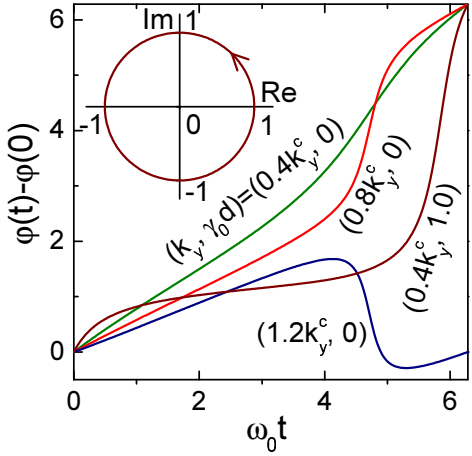}
\caption{Argument of the complex reflection amplitude,
$\varphi(t)=\mbox{arg}(r_{\uparrow\uparrow})$,  as a function
of $\omega_{0}t$ for four sets of $(k_{y},\gamma_{0}d)$.
The other parameters are taken to be $R_{0}=0$ and
$v_{\mbox{\tiny F}}eA_{0}=M_{0}$ with $k_{y}^{c}=e\vert A_{0}\vert$.
Inset: trajectories of $r_{\uparrow\uparrow}$
in a cycle on the complex plane.}
\label{Fig_2}
\end{figure}
We first consider  the case of $R_{0}=0$. From Eq.\ (\ref{Rupup0}),
it is easy to show $\vert r_{\uparrow\uparrow}(k_y)\vert=1$. In Fig.\ \ref{Fig_2},
we plot the argument  $\varphi(t)$  of $r_{\uparrow\uparrow}(k_y)$
 as a function of $\omega_{0}t$ for several parameter sets.
For either $\gamma_{0}d=0$ (ideal contact) or $1.0$ (strong potential barrier),
$\varphi(t)$ always increments $2\pi$ in a cycle
as long as $\vert k_{y}\vert<k_{y}^{c}$. In this case,
the trajectories of $r_{\uparrow\uparrow}(k_y)$ always form
a unit circle on the complex plane, oriented counterclockwise, as shown
in the inset of Fig.\ \ref{Fig_2},
suggesting $\Delta s_{z}(k_y)=\hbar$ (for $E_{0}M_{0}>0$).
For $\vert k_y\vert>k_{y}^{c}$, however, the situation is quite different.
$\varphi(t)$ does not change after going through a cycle,
and the trajectory of $r_{\uparrow\uparrow}(k_y)$
does not enclose a finite area, so that  $\Delta s_{z}(k_y)=0$.
Apparently,
the present result conforms to the
spin Chern numbers given by Eq.\ (\ref{sChern})
 and the spectral flows of the SPWFs.

\begin{figure}
\includegraphics[width=2.6in]{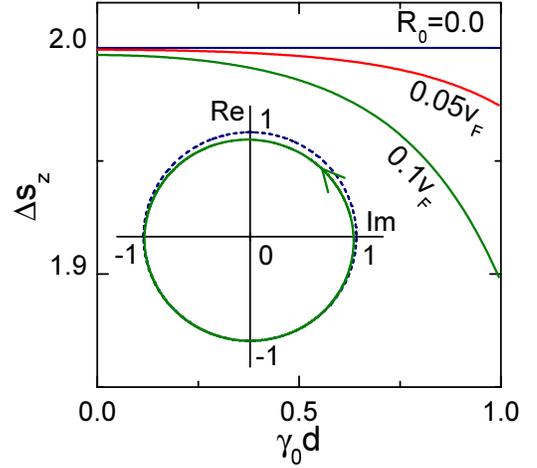}
\caption{$\Delta s_{z}(k_y)$ (in unit of $\hbar/2$)
as a function of
$\gamma_{0}d$ for $k_{y}=0.4k_{y}^{c}$ and three
different values of $R_{0}$. The other parameters are taken to be
the same as in Fig.\ \ref{Fig_2}.
Inset: the trajectory of $r_{\uparrow\uparrow}$ in a cycle
for $R_{0}=0.1v_{\mbox{\tiny F}}$
and $\gamma_{0}d=1.0$, with the unit circle indicated by
the dotted line.} \label{Fig_3}
\end{figure}
Next we study the correction to $\Delta s_{z}(k_y)$
due to nonzero Rashba spin-orbit coupling.
By expressing $r_{\uparrow\uparrow}=\rho e^{i\varphi}=
\sqrt{1-\delta\rho^2}e^{i(\varphi^{(0)}+\delta\varphi)}$
in the polar coordinate system,
where $\varphi^{(0)}$  is the argument at $R_0=0$, and
$\delta\varphi$ and $\delta\rho^2$ stand for the
second-order corrections to $\varphi$ and $\rho^2$, respectively,
due to the Rashba spin-orbit coupling, Eq.\ (\ref{DS2})
becomes $\Delta s_{z}(k_y)=(\hbar/2\pi)[\oint_{T}
(1-\delta\rho^2)d\varphi^{(0)}+\oint_{T}d\delta\varphi]
+{\cal O}(\epsilon^3)$. We notice that
$\delta\varphi$ is a small quantity fluctuating around
$0$ and periodic in time,
$\delta\varphi\vert_{t=0}=\delta\varphi\vert_{t=T}$,
so that $\oint_{T}d\delta\varphi=0$. Using the identity
$\delta\rho^2=\vert r_{\downarrow\uparrow}\vert^2$, we then obtain
$\Delta s_{z}(k_y)=(\hbar/2\pi)\oint_{T}
(1-\vert r_{\downarrow\uparrow}\vert^2)d\varphi^{(0)}$,
where $d\varphi^{(0)}$ can be calculated from Eq.\ (\ref{Rupup0})
and $r_{\downarrow\uparrow}$ has been given by Eq.\ (\ref{Rdnup}).
This is an expression for $\Delta s_{z}(k_y)$
accurate to the second order
in $R_{0}/v_{\mbox{\tiny F}}$.
At $\gamma_{0}d=0$, we  obtain $\Delta s_{z}(k_y)$
as
\begin{equation}
\Delta s_{z}(k_y)\simeq\frac{\hbar}{2}\left[
1-\frac{5}{32}\left(\frac{R_{0}}{v_{\mbox{\tiny F}}}\right)^{2}
\right](C_{+}-C_{-})\ ,
\label{DS_R}
\end{equation}
for $\vert k_y\vert<k_{y}^c$, and $\Delta s_{z}(k_y)=0$
elsewhere. As expected, nonzero Rashba spin-orbit coupling
causes $\Delta s_{z}(k_y)$ to deviate
from its quantized value, i.e., $(C_{+}-C_{-})\hbar/2$.
In real materials, $R_{0}$ is usually much
smaller (by an order of magnitude or more) than $v_{\mbox{\tiny F}}$,~\cite{HgTeRashba}
so that  the deviation is
less than a percent for an ideal connection
between the pump and electrode.

For $\gamma_{0}d\neq 0$, $\Delta s_{z}(k_y)$
can be evaluated numerically and its calculated result is plotted in Fig.\ \ref{Fig_3} as a
function of $\gamma_{0}d$ for three different strengths
of the Rashba spin-orbit coupling. For $R_{0}=0$,
$\Delta s_{z}(k_y)$ is quantized to $\hbar$, independent of
$\gamma_{0}d$. For $R_{0}=0.05v_{\mbox{\tiny F}}$ and $0.1v_{\mbox{\tiny F}}$,
weak potential barrier ($\gamma_{0}d\ll 1$) has little
effect on $\Delta s_{z}(k_y)$. This is reasonable
as the leading-order correction of small $\gamma_{0}d$
must be ${\cal O}(\epsilon^2\gamma_{0}d)$.
Appreciable deviations from the quantized value
occur for strong potential barrier (e.g., $\gamma_{0}d\simeq 1$).
We note that $\delta\varphi$ does not affect the orbit
of $r_{\uparrow\uparrow}$,
as it represents a variation in the tangent direction
of the orbit, and the orbit can be
determined by $r_{\uparrow\uparrow}=
\sqrt{1-\vert r_{\downarrow\uparrow}\vert^2}e^{i\varphi^{(0)}}$.
Inset shows the
trajectory of $r_{\uparrow\uparrow}$ on the complex
plane for $R_{0}=0.1v_{\mbox{\tiny F}}$ and $\gamma_{0}d=1.0$.
For such a strong potential barrier,
the orbit of $r_{\uparrow\uparrow}$
deviates from the unit circle visibly. The above result
suggests that improving the contact quality between
the pump and electrode is helpful for obtaining a nearly
integer-quantized value of the pumped spin.

By summing over $k_y$ between $-k_{y}^{c}$
and $k_{y}^{c}$, we obtain for the total spin pumped per cycle
\begin{equation}
\Delta S_{z}=\sigma_{s}(2\vert E_{0}\vert L_{y}/\omega_{0})\ ,
\end{equation}
with
\begin{equation}
\sigma_{s}\simeq \frac{e}{4\pi}\left[
1-\frac{5}{32}\left(\frac{R_{0}}{v_{\mbox{\tiny F}}}\right)^{2}\right]
(C_{+}-C_{-})\ ,
\end{equation}
for a good contact ($\gamma_{0}d\ll 1$).
$\Delta S_{z}$ is in scale with
width $L_{y}$ of the pump. By noting that $\omega_{0}$ is
proportional to the number of cycles per unit time, $\sigma_{s}$ can be considered
as the spin pumping conductivity.

\subsection{Charge pumping for a half-metallic electrode}
Now we discuss a possible way to experimentally observe
the spin Chern numbers, by using a half-metallic electrode,
in which conducting channels for electron spin antiparallel
to the spin polarization
are absent. We first consider the case, where
the spin polarization of the electrode is parallel to
the $z$ axis. The Hamiltonian
of the electrode is taken as $H_{E}=v'_{\mbox{\tiny F}}
(k_{y})k_x\hat{s}_{z}\hat{\sigma}_{x}+
V_{1}(\hat{1}-\hat{s}_{z})\hat{\sigma}_{z}/2$.
In this case, as shown in Appendix C,
$r_{\uparrow\uparrow}$ is still given by Eq.\ (\ref{Rupup0})
but $r_{\downarrow\uparrow}\equiv 0$. It follows that
for any $k_{y}$ between $ -k_{y}^{c}$ and $ k_{y}^{c}$, the
charge pumped per cycle is integer-quantized and equal to
 $\Delta q(k_y)=(-e)C_{+}$. Similarly, for the spin
polarization of the electrode antiparallel to the $z$ axis,
the charge pumped is equal to $\Delta q(k_y)=(-e)C_{-}$. Therefore, the total
charge pumped per cycle is given by
\begin{equation}
\Delta Q=\sigma_{c}(2\vert E_{0}\vert L_y/\omega_{0})\ ,
\label{totCharge}
\end{equation}
with
\begin{equation}
\sigma_{c}
=-C_{\pm}\frac{e^2}{h}\ ,
\label{cCond}
\end{equation}
where the spin Chern number $C_{+}$ ($C_{-}$) is taken
for the spin polarization of the electrode
parallel (antiparallel) to the $z$ axis.
We emphasize that Eqs.\ (\ref{totCharge}) and (\ref{cCond}) obtained above
are valid for finite Rashba spin-orbit coupling and finite potential barrier
between the pump and electrode,
indicating that the quantized
charge pumping is robust against small perturbations.
Experimentally, $\Delta Q$ can be obtained by measuring the electrical
current in the electrode, and from Eqs.\ (\ref{totCharge}) and (13),
$C_{\pm}$  can be evaluated,  yielding  an experimental method to measure
the spin Chern numbers directly. The sign inversion
of $\Delta Q$ with reversing the spin polarization of the
half-metallic electrode, as indicated by Eqs.\ (\ref{totCharge}) and (\ref{cCond}),
is a hallmark of the present spin Chern
charge pump, which can be used to distinguish it
from the conventional Thouless charge pump~\cite{Thouless1,Thouless2}. 

We have used the single-electron 
approximation, where the electron interaction 
is not taken into account. In particular, in the half-metallic 
electrode case, one of the spin channel is blocked at the 
boundary, which naturally induces some charge and spin 
accumulations, and consequently changes the potential profile. However, 
owing to the screening effect, the change in the potential profile 
is expected to be localized at the boundary, which in effect
modifies the potential barrier between the 
pump body and the electrode. As has been shown above, the charge pumping
effect is independent of the existence and details of the potential barrier, and
so we believe that the pumping effect will survive the charge and spin 
accumulations.   \\
\\
\section{Discussion}
Up to now, all the results
obtained from the scattering matrix formula are apparently
in complete agreement with the spin Chern numbers
given by Eq.\ (\ref{sChern}). These results cannot be
explained within the framework of the $Z_2$ theory.~\cite{Z2pump}
While one can define a $Z_2$ index at the TR-invariant point $k_y=0$, the
effective 1D Hamiltonian  given by
Eqs.\ (\ref{Hamil}) and (\ref{Rashba}) for any given nonzero $k_{y}$
does not preserve the TR symmetry, as its TR partner
is at $-k_{y}$, making the $Z_{2}$ index invalid. The $Z_{2}$ theory
predicted that the TR symmetry is crucial for the topological spin pumping,~\cite{Z2pump}
suggesting that only the states at the TR-invariant point $k_y=0$
can contribute to the spin pumping. This clearly contradicts the present
result that all the states with $\vert k_y\vert<k_{y}^{c}$  contribute equally,
which is obtained directly from the scattering matrix formula.
This point is also evidenced
by the fact that the total amount of charge or spin pumped
per cycle is in proportion to the sample width $L_{y}$.
For the same reason, the pumping effect
found in this
work is also essentially different from that via
edge states in Refs.~\cite{RealizeZ2_0,RealizeZ2_1}, where the amount of spin
or charge pumped per cycle is proportional to the number of the
gapless edge channels.

In conclusion, our work uncovers a bulk topological pumping effect
due to direct transfer of the SPWFs between the pump and electrode,
without the participation of edge states. This measurable effect reveals
the bulk topological properties of the system
that are neither captured by the $Z_2$ index nor
reflected by the number of gapless edge channels.
It can be accurately
described by the spin Chern numbers. This spin
Chern pump may lay the foundation for
direct experimental study and possibly utilization of the
robust topological properties of the TIs. 

The previous experimental 
work~\cite{SciExp} evidenced the difficulty of modulating in time
the properties of an open quantum dot without generating undesired bias voltages due
to stray capacitances. This problem might not be significant 
in our pumping setup, where a much larger
bulk sample of the TI can be used and the stray capacitances can be greatly
reduced. Moreover, a possible way around the
obstacle is to use the $ac$ Josephson effect to induce periodically time-dependent
Andreev reflection amplitudes in a hybrid 
normal-superconducting system.~\cite{NatPhysExp} Concrete design of a spin Chern pump
based upon the Josephson effect will await future work.
While the proposed spin pumping scheme may have the advantage of low noises, 
its practical application in spintronic devices still relies on 
the discovery of new TIs with bulk band gaps much greater than
room temperature, which determine the temperature range where
the spin Chern pumping effect can survive. Currently, precessing magnetization
is a feasible method to generate robust spin currents in spintronic
devices at room temperature.~\cite{PhysicsExp}

%Our approach is
%different from the previously suggested methods~\cite{RealizeZ2_0,RealizeZ2_1}
%for measuring the bulk topological invariants, where the
%spin or charge is pumped through the symmetry-protected helical
%edge states of the QSH systems. These latter methods directly probe
%the properties of the edge states, where the amount of spin
%or charge pumped per cycle is proportional to the number of gapless edge
%channels. In contrast, the spin or charge pumping effect
%discovered in this work is a direct transfer of the bulk
%spin-polarized Wannier functions
%between the pump and electrode, without the participation
%of edge states. (The sample width was
%taken to be sufficiently large,
%such that any edge effects were neglected with respect to
%the bulk contribution.)

%In this sense, our method reveals the topological invariants,
%the spin Chern numbers, more directly from the bulk
%point of view.
%In contrast,
%the spin Chern numbers
%have accurately described the topological pumping effect
%in the present system.

\begin{acknowledgments}
This work was supported by the State Key Program for Basic Researches of China under
grants numbers 2015CB921202, 2014CB921103 (LS), 2011CB922103 and 2010CB923400 (DYX), the National
Natural Science Foundation of China under grant numbers 11225420 (LS),
11174125, 91021003 (DYX) and a project funded by the PAPD of Jiangsu Higher
Education Institutions. We also thank the US NSF grants numbers DMR-0906816 and
DMR-1205734 (DNS).
\end{acknowledgments}

%\section{Methods}
\appendix

\section{A POSSIBLE EXPERIMENTAL REALIZATION OF THE SPIN CHERN PUMP}

In what follows we expand on the model setup and
possible experimental realization of the spin Chern pump
in more details.

\subsection{The pump}
\begin{figure}
\includegraphics[width=1.7in]{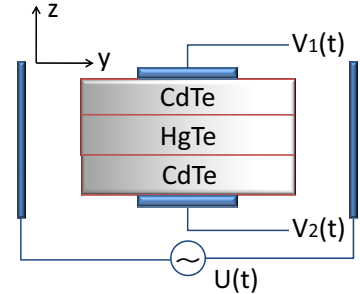}
\caption{
A schematic view of a experimental setup for realization
of a 2D spin Chern pump. A CdTe/HgTe/CdTe quantum well
heterostructure, with dual gates on its top and bottom,
is placed between two conductive plates. When the voltages
of the gates and plates are adiabatically modulated in proper
cycles, spin or charge can be pumped into electrodes
coupled to the quantum well
along the $x$ direction.} \label{Sch}
  \end{figure}

A possible experimental realization of the Hamiltonian
Eq.\ (\ref{Hamil}) for the pump is illustrated
in Fig.\ \ref{Sch}. A HgTe/CdTe quantum-well heterostructure with dual gates (top
and bottom) is placed between two conductive plates. It is known that
when the width of the quantum well (thickness of the HgTe film)
is above a critical size $d_{c}=6.3nm$,~\cite{BHZModel,BHZModel1}
the band structure is inverted,
characterized a negative mass term $-M_{0}\hat{\sigma}_{z}$ in the Hamiltonian,
corresponding to the QSH state.
If the width of the quantum well falls below $d_{c}$, the band structure
will be aligned in a ``normal'' way with a
positive mass term $M_{0}\hat{\sigma}_{z}$,
corresponding to a normal insulator. As has been discussed
in Refs.\ \cite{DualGate1} and \cite{DualGate2}, the
topological phase transition between the QSH phase and normal insulator can
also be tuned by applying a gate voltage, which effectively
reduces the width of the quantum well.
It is assumed that the quantum well under consideration
has a width somewhat greater than $d_{c}$, and so
has a negative mass term $-M_{0}\hat{\sigma}_{z}$ initially.
With increasing the gate voltage, the electron mass
increases, and can invert its sign. Usually,
increasing the gate voltage may
also adjust the carrier density. Nevertheless, it has
been shown~\cite{DualGate1,DualGate2} that, by
using dual gates and properly tuning their voltages
$V_{1}(t)$ and $V_{2}(t)$, it is generally possible to
change the electron mass in the desired manner to be
$-M_{0}\cos(\omega_{0}t)$, while keeping
the electron Fermi energy still in the band gap.

The effect of the conductive plates
is easily  understood. When a voltage drop $U(t)$ is applied
across the plates,
a uniform electric field ${\bf E}(t)=E(t)\hat{\bf y}$
will be generated in the space between the two plates.
The electrons in the quantum well
experience a vector potential ${\bf A}(t)=A(t)\hat{\bf y}$
with $A(t)$ defined as $E(t)=-\partial A(t)/\partial t$.
If the electric field is chosen to be $E(t)=-E_{0}\cos(\omega_{0}t)$,
one gets $A(t)=A_{0}\sin(\omega_{0}t)$ with $A_{0}=E_{0}/\omega_{0}$,
as desired. We point out that the exact time dependencies
of $M(t)$ and $A(t)$ are  not essential for realizing
the spin Chern pump, provided that they have the same
periodicity and a constant relative phase shift.
\\

\subsection{The nonmagnetic electrode}
The pumping effect is
insensitive to material details of the electrode.
The electrode is taken to be
a normal metal with a 2D parabolic Hamiltonian
$H_{E}=-E_{0}+p^2/2m$.
When $E_{0}$ is sufficiently large,
for a given $p_{y}$, we can linearize the effective
1D Hamiltonian $H_{E}$ at the right and left
Fermi points $p_{x}=\pm mv'_{\mbox{\tiny F}}(k_y)$
with $v'_{\mbox{\tiny F}}(k_y)=\sqrt{2m(E_{\mbox{\tiny F}}+E_{0})-k_{y}^2}/m$.
A Pauli matrix $\hat{\sigma}_x$ is introduced to
describe the two branches. To be consistent with the
form of the Hamiltonian in the pump,  we use
$\sigma_{x}=1$ and $-1$,  respectively,
to represent the right-moving and left-moving
branches for $s_{z}=1$ and oppositely for $s_{z}=-1$.
As a result, the Hamiltonian of the electrode becomes
$H_{E}=v'_{\mbox{\tiny F}}(k_{y})k_x\hat{s}_{z}\hat{\sigma}_{x}$
at $E_{\mbox{\tiny F}}=0$, where $k_y=p_{y}$
and $k_x=p_{x}\mp mv'_{\mbox{\tiny F}}(k_y)$.
The spin pumping effect is usually dominated
by small $k_y$, so that we can further approximate
$v'_{\mbox{\tiny F}}(k_y)\simeq v'_{\mbox{\tiny F}}(k_{y}=0)\equiv
v'_{\mbox{\tiny F}}$, with purpose to minimize the number
of adjustable parameters in the model.

\subsection{The barrier}
For the present Dirac-like Hamiltonian,
an ordinary potential barrier has a very weak effect on
the electron transmission due to the Klein paradox.
Therefore, we take the Hamiltonian for the barrier
to be $H_{B}=H_{E}+V_{0}\hat{\sigma}_z$.
The inclusion of potential $V_{0}\hat{\sigma}_z$ opens up an insulating
energy gap of size $2V_{0}$ around the Fermi level, which
 presumably is more efficient for describing the
contact deficiencies and structural mismatch between
the pump and electrode.

\subsection{The half-metallic electrode}
The half metal, e.g., CrO$_{2}$, La$_{2/3}$Sr$_{1/3}$MnO$_{3}$, etc., is a substance
that acts as a conductor to electrons of one spin orientation,
but as an insulator to those of the other
spin orientation.  From the viewpoint of the electronic structure,
one of the spin subbands is metallic, whereas the Fermi
level falls into an energy gap of the other spin subband.
To simulate the half-metallic electrode, $H_{E}$ is taken to
be $H_{E}=v'_{\mbox{\tiny F}}(k_{y})k_x\hat{s}_{z}\hat{\sigma}_{x}+
V_{1}(\hat{1}\mp\hat{s}_{z})\hat{\sigma}_{z}/2$, where $\mp$ stands for
the spin polarization of the electrode parallel and antiparallel to the $z$-axis,
respectively. The second term opens an energy gap
of size $2V_{1}$ around the Fermi level
for electron spin antiparallel to the spin polarization
of the electrode, without affecting the other spin subband.
As a result, the electron density of states is fully spin-polarized
at the Fermi energy. $V_{1}$ is set to be infinity in the final result.

%\appendix

\section{CALCULATION OF THE REFLECTION AMPLITUDES FOR A NONMAGNETIC ELECTRODE}

\subsection{Electron wavefunctions in the pump and potential barrier}
We now solve the scattering problem for an electron at the Fermi energy
incident from the electrode. The Fermi energy will be taken to be $E_{\mbox{\tiny F}}=0$,
which is in the band gap of the pump. Therefore, the incident electron
will be fully reflected back into the electrode.
The Rashba spin-orbit
coupling is treated as a perturbation, and the result
will be calculated to the linear order in the small quantity $\epsilon=R_{0}/
v_{\mbox{\tiny F}}$. The wavefunctions of the pump ($x<0$), barrier ($0<x<d$) and electrode
($x>d$) are denoted by $\Psi_{P}(x)$, $\Psi_{B}(x)$, and $\Psi_{E}(x)$, respectively.
We have two boundary conditions: $\Psi_{P}(0^{-})=\Psi_{B}(0^{+})$
and $\Psi_{B}(d-0^{+})=\Psi_{E}(d+0^{+})$.

We use $\uparrow$ and $\downarrow$
to represent the eigenstates of $\hat{s}_{z}$, and $+1$ and $-1$ to
represent those of $\hat{\sigma}_{z}$. On the basis $\vert
\uparrow,+1\rangle$, $\vert\uparrow,-1\rangle$, $\vert
\downarrow,+1\rangle$, and $\vert\downarrow,-1\rangle$,
the Hamiltonian of pump (the Eqs.\ (1) and (2) in the manuscript) can be
expanded as a $4\times 4$ matrix
\begin{widetext}
\begin{equation}
H_{P}=\left(
\begin{array}{cccc}
-M(t)&v_{\mbox{\tiny F}}(k_{x}+i\tilde{k}_{y})&R_{0}(-ik_{x}-\tilde{k}_{y})&0\\
v_{\mbox{\tiny F}}(k_{x}-i\tilde{k}_{y})&M(t)&0&0\\
R_{0}(ik_{x}-\tilde{k}_{y})&0&-M(t)&v_{\mbox{\tiny F}}(-k_{x}+i\tilde{k}_{y})\\
0&0&v_{\mbox{\tiny F}}(-k_{x}-i\tilde{k}_{y})&M(t)
\end{array}
\right)
\label{Hp_mat}
\end{equation}
\end{widetext}
where $\tilde{k}_{y}=k_{y}+eA(t)$. For energy $E=E_{\mbox{\tiny F}}=0$,
the eigen-equation is obtained from Eq.\ (\ref{Hp_mat})
\begin{equation}
[M^2(t)+v^2_{\mbox{\tiny F}}\tilde{k}^{2}]^2-M^2(t)R^2_{0}\tilde{k}^2=0\ ,
\label{eigen_mat}
\end{equation}
with $\tilde{k}^2=k_{x}^2+\tilde{k}_{y}^2$,
and up to a normalization factor, the eigenfunctions are
\begin{equation}
\left(
\begin{array}{c}
A_{1}M(t)/v_{\mbox{\tiny F}}\\
-A_{1}(k_{x}-i\tilde{k}_{y})\\
A_{2}M(t)/v_{\mbox{\tiny F}}\\
A_{2}(k_{x}+i\tilde{k}_{y})
\end{array}
\right)e^{ik_{x}x/\hbar}\ ,
\label{eigenFunc}
\end{equation}
where $A_{1}=-(ik_{x}+\tilde{k}_{y})M(t)$ and $A_{2}=[M^2(t)+v_{\mbox{\tiny F}}^2
\tilde{k}^2]/R_{0}$. We need to solve $k_x$ from the eigen-equation Eq.\ (\ref{eigen_mat}).
We notice that the equation is a 4th-degree polynomial of $k_x$
with real coefficients, so complex conjugate roots must appear in pairs. Moreover,
Eq.\ (\ref{eigen_mat}) is even in $k_x$, so positive and negative roots appear in pairs.
In combination, Eq.\ (\ref{eigen_mat}) must have four roots of the form
$k_{x}=a+ib$, $a-ib$, $-a+ib$, and $-a-ib$. By substitution of the four roots
into Eq.\ (\ref{eigenFunc}), we can in principle obtain four different eigenfunctions.
For the present scattering problem, we only need the two eigenfunctions that are
decaying into the pump, which correspond to the two roots with negative imaginary
parts.

For $R_{0}=0$, it is easy to obtain for the roots
for the two decaying modes:
$k_{x}=-i\hbar\eta$ with $\hbar\eta=\sqrt{M^2(t)+
\tilde{k}_{y}^2}$,
which are two-fold degenerate. The corresponding
two decaying eigenfunctions are given by
\begin{equation}
\varphi_{+}(x)=\vert \uparrow\rangle\otimes\left(
\begin{array}{c}
\sin\theta\\
i\cos\theta
\end{array}
\right)e^{\eta x}\ ,
\label{eigenFuncWithoutR00}
\end{equation}
\begin{equation}
\varphi_{-}(x)=\vert \downarrow\rangle\otimes\left(
\begin{array}{c}
\cos\theta\\
-i\sin\theta
\end{array}
\right)e^{\eta x}\ ,
\label{eigenFuncWithoutR01}
\end{equation}
where $2\theta=\mbox{Arg}[v_{\mbox{\tiny F}}\tilde{k}_{y}+iM(t)]$. For $R_{0}\neq 0$,
we write the roots of $k_{x}$ as $k_{x}=-i\hbar\eta+\delta k_{x}$,
and also write $k_{x}^2$ as $k_{x}^2=-(\hbar\eta)^2+\delta k_{x}^2$.
To the second order in $\epsilon$, we can solve for $\delta k_{x}^2$
from the eigen-equation Eq.\ (\ref{eigen_mat})
\begin{equation}
\delta k_{x}^2=\left(
\pm i\frac{R_{0}}{v_{\mbox{\tiny F}}}+\frac{R^2_{0}}{2v^2_{\mbox{\tiny F}}}
\right)(\hbar\eta)^2\sin^2(2\theta)+{\cal O}(\epsilon^3)\ .
\label{dKx2}
\end{equation}
Noticing that the expression for $A_{2}$
given below Eq.\ (\ref{eigenFunc}) has a factor $R_{0}$ in the denominator,
we keep $\delta k_{x}^2$ to the second order, for the purpose to
calculate $A_{2}$ to the linear order. By using the relation $\delta k_{x}^2=
-2i\hbar\eta\delta k_{x}+{\cal O}(\epsilon^2)$, we derive from Eq.\ (\ref{dKx2})
\begin{equation}
\delta k_{x}=\mp\frac{R_{0}}{2v_{\mbox{\tiny F}}}\hbar\eta\sin^2(2\theta)
+{\cal O}(\epsilon^2)\ .
\end{equation}
With these relations, we obtain for $A_{1}$ and $A_{2}$
\begin{eqnarray}
A_{1}&=&-v_{\mbox{\tiny F}}(\hbar\eta)^{2}\sin(2\theta)[1+\cos(2\theta)]\\
&\pm&i\frac{R_{0}}{2}(\hbar\eta)^2\sin^3(2\theta)+{\cal O}(\epsilon^2)\ ,
\end{eqnarray}
and
\begin{equation}
A_{2}=\left(\pm i+\frac{R_{0}}{2v_{\mbox{\tiny F}}}\right)v_{\mbox{\tiny F}}(\hbar\eta)^2
\sin^2(2\theta)+{\cal O}(\epsilon^2)\ .
\end{equation}
We can always eliminate any common factor that appears in all
the four components of Eq.\ (\ref{eigenFunc}), whenever possible.
By eliminating a common factor $2v_{\mbox{\tiny F}}(\hbar\eta)^2\sin(2\theta)\cos\theta$,
we rewrite $A_{1}$ and $A_{2}$ as
\begin{equation}
A_{1}=-\cos\theta
\pm i\frac{R_{0}}{2v_{\mbox{\tiny F}}}\sin\theta\sin(2\theta)+{\cal O}(\epsilon^2)\ ,
\end{equation}
and
\begin{equation}
A_{2}=\left(\pm i+\frac{R_{0}}{2v_{\mbox{\tiny F}}}\right)
\sin\theta+{\cal O}(\epsilon^2)\ .
\end{equation}
Then the two decaying wavefunctions can be derived to be
\begin{equation}
\varphi_{1,2}(x)=
%\left(
%\begin{array}{c}
%A_{1}\cos\theta\sin\theta\\
%A_{1}i\cos^2\theta(1\mp i\frac{R_{0}}{v_{\mbox{\tiny F}}}
%\sin^2\theta)\\
%A_{2}\sin\theta\cos\theta\\
%-A_{2}i\sin^2\theta(1\mp i\frac{R_{0}}{v_{\mbox{\tiny F}}}\cos^2\theta)
%\end{array}
%\right)e^{\eta_{\mp}x}\\
%=
\left(
\begin{array}{c}
-\cos^2\theta\sin\theta\left(1\mp i\frac{R_{0}}{v_{\mbox{\tiny F}}}\sin^2\theta\right)\\
-i\cos^3\theta\left(1\mp 2i\frac{R_{0}}{v_{\mbox{\tiny F}}}\sin^2\theta\right)\\
\pm i\sin^2\theta\cos\theta\left(1\mp i\frac{R_{0}}{2v_{\mbox{\tiny F}}}\right)\\
\pm \sin^3\theta\left[1\mp i\left(\frac{1}{2}+\cos^2\theta\right)\frac{R_{0}}{v_{\mbox{\tiny F}}}\right]
\end{array}
\right)e^{\eta_{\mp}x}\ ,
\label{eigenFuncR0}
\end{equation}
where $\eta_{\mp}=\eta\bigl[1\mp\frac{R_{0}}{2v_{\mbox{\tiny F}}}\sin^2(2\theta)\bigr]$.

Some remarks are in order. With respect to
the wavefunctions at $R_{0}=0$, namely, Eqs.\ (\ref{eigenFuncWithoutR00})
and (\ref{eigenFuncWithoutR01}),
nonzero $R_{0}$ leads to nonperturbative change in the
wavefunctions Eq.\ (\ref{eigenFuncR0}),
in the sense that Eq.\ (\ref{eigenFuncR0}) will not recover
Eqs.\ (\ref{eigenFuncWithoutR00})
and (\ref{eigenFuncWithoutR01}) in the limit $R_{0}\rightarrow 0$.
This is reasonable, just as what always happens in
the degenerate perturbation theory of the quantum mechanics.
For the present problem,
the wavefunction $\Psi_{P}(x)$ in the pump is always expressed as an arbitrary
linear superposition of the two decaying modes:
$\Psi_{P}(x)=B_{1}\varphi_{1}(x)+B_{2}\varphi_{2}(x)$.
By defining $\varphi_{+}(x)=-[\varphi_{1}(x)+\varphi_{2}(x)]$
and $\varphi_{-}(x)=-i[\varphi_{1}(x)-\varphi_{2}(x)]$,
we can rewrite $\Psi_{P}(x)$ as $\Psi_{P}(x)=D_{1}\varphi_{+}(x)+D_{2}\varphi_{-}(x)$.
The final result for the reflection amplitudes
depends only on $\Psi_{P}(x=0^{-})$,
so we explicitly write out the expression for $\Psi_{P}(x=0^{-})$
as follows
\begin{equation}
\Psi_{P}(x=0^{-})=D_{1}\varphi_{+}(0^{-})+D_{2}\varphi_{-}(0^{-})\ ,
\label{PsiM}
\end{equation}
where
\begin{equation}
\varphi_{+}(0^{-})=\left(
\begin{array}{c}
\sin\theta\\
i\cos\theta\\
-\frac{\sin^2\theta}{\cos^2\theta}\cos\theta\frac{R_{0}}{2v_{\mbox{\tiny F}}}\\
i\left(\frac{1}{2}+\cos^2\theta\right)\frac{\sin^3\theta}{\cos^2\theta}
\frac{R_{0}}{v_{\mbox{\tiny F}}}
\end{array}
\right)\ ,
\end{equation}
\begin{equation}
\varphi_{-}(0^{-})=\left(
\begin{array}{c}
\cos^2\theta\sin\theta\frac{R_{0}}{v_{\mbox{\tiny F}}}\\
2i\cos^3\theta\frac{R_{0}}{v_{\mbox{\tiny F}}}\\
\cos\theta\\
-i\sin\theta
\end{array}
\right)
\end{equation}
Now we see that in the limit $R_{0}\rightarrow 0$,
the total wavefunction $\Psi_{P}(0^{-})$ will go back to the form
of a superposition of the two decaying wavefunctions given in
Eqs.\ (\ref{eigenFuncWithoutR00})
and (\ref{eigenFuncWithoutR01}). In conclusion, while
small Rashba spin-orbit coupling may cause a nonperturbative
change of the individual decaying wavefunctions, it modifies the ``space''
spanned by the two decaying modes in a perturbative manner.
It is this ``space'' which determines the final result of
the reflection amplitudes. This is the physical reason why in
the final result, the Rashba spin-orbit coupling modifies
the reflection amplitudes in a perturbative manner.
The wavefunction in the potential barrier can be written as
\begin{widetext}
\begin{eqnarray}
\Psi_{B}(x)=\frac{C_{1}}{\sqrt{2}}\vert\uparrow\rangle\otimes\left(
\begin{array}{c}
1\\
-i
\end{array}
\right)
e^{\gamma_{0}x}+\frac{C_{2}}{\sqrt{2}}\vert\uparrow\rangle\otimes\left(
\begin{array}{c}
1\\
i
\end{array}
\right)
e^{-\gamma_{0}x}+
\frac{C_{3}}{\sqrt{2}}\vert\downarrow\rangle\otimes\left(
\begin{array}{c}
1\\
i
\end{array}
\right)
e^{\gamma_{0}x}+\frac{C_{4}}{\sqrt{2}}\vert\downarrow\rangle\otimes\left(
\begin{array}{c}
1\\
-i
\end{array}
\right)
e^{-\gamma_{0}x}\ ,
\label{WaveBarrier}
\end{eqnarray}
\end{widetext}
where $\gamma_{0}=V_{0}/\hbar v'_{\mbox{\tiny F}}$.

\subsection{An electron incident from the spin-up channel}
For an electron incident from the spin-up channel, the wavefunction in the electrode
is given by
\begin{eqnarray}
\Psi_{E}(x)
&=&\frac{1}{\sqrt{2}}\vert\uparrow\rangle\otimes\left(
\begin{array}{c}
1\\
-1
\end{array}
\right)+\frac{r_{\uparrow\uparrow}}{\sqrt{2}}\vert\uparrow\rangle\otimes\left(
\begin{array}{c}
1\\
1
\end{array}
\right)\nonumber\\
&+&\frac{r_{\downarrow\uparrow}}{\sqrt{2}}\vert\downarrow\rangle\otimes\left(
\begin{array}{c}
1\\
-1
\end{array}
\right)\ .
\label{Waveelectrode}
\end{eqnarray}
First, matching the wavefunctions Eqs.\ (\ref{WaveBarrier}) and (\ref{Waveelectrode})
at $x=d$, one obtain
\begin{eqnarray}
C_{1}&=&\frac{1}{2}[(1-i)+r_{\uparrow\uparrow}(1+i)]e^{-\gamma_{0}d}\ ,
\label{C_1}\\
C_{2}&=&\frac{1}{2}[(1+i)+r_{\uparrow\uparrow}(1-i)]e^{\gamma_{0}d}\ ,
\label{C_2}\\
C_{3}&=&\frac{r_{\downarrow\uparrow}}{2}(1+i)e^{-\gamma_{0}d}\ ,
\label{C_3}\\
C_{4}&=&\frac{r_{\downarrow\uparrow}}{2}(1-i)e^{\gamma_{0}d}\ .
\label{C_4}
\end{eqnarray}
In the next step, we will match wavefunctions at $x=0$. Substituting
Eqs.\ (\ref{C_1}-\ref{C_4}) into Eq.\ (\ref{WaveBarrier}), we can write
Eq.\ (\ref{WaveBarrier}) at $x=0^{+}$ as
\begin{equation}
\Psi_{B}(0^{+})
=\frac{1}{\sqrt{2}}
\left(
\begin{array}{c}
\Gamma_{+}(2\gamma_{0}d)+r_{\uparrow\uparrow}\Gamma_{-}(2\gamma_{0}d)\\
-\Gamma_{-}(2\gamma_{0}d)+r_{\uparrow\uparrow}\Gamma_{+}(2\gamma_{0}d)\\
r_{\downarrow\uparrow}\Gamma_{-}(2\gamma_{0}d)\\
-r_{\downarrow\uparrow}\Gamma_{+}(2\gamma_{0}d)
\end{array}
\right)
\label{PsiP0}
\end{equation}
where $\Gamma(\xi)=\mbox{ch}(\xi)\pm i\mbox{sh}(\xi)$. Now
equating Eq.\ (\ref{PsiM}) with Eq.\ (\ref{PsiP0}), we obtain
\begin{eqnarray}
&&\left(\begin{array}{cc}
i\sin^2\theta&\sin\theta\cos\theta\\
-\sin\theta\cos\theta&i\cos^2\theta
\end{array}\right)
\left(
\begin{array}{c}
\Gamma_{-}\\
-\Gamma_{+}
\end{array}
\right)
\frac{r_{\downarrow\uparrow}}
{i\frac{R_{0}}{v_{\mbox{\tiny F}}}\cos\theta\sin^3\theta}\nonumber\\
&=&\left(
\begin{array}{c}
\Gamma_{+}(2\gamma_{0}d)+r_{\uparrow\uparrow}\Gamma_{-}(2\gamma_{0}d)\\
-\Gamma_{-}(2\gamma_{0}d)+r_{\uparrow\uparrow}\Gamma_{+}(2\gamma_{0}d)
\end{array}
\right)+{\cal O}(\epsilon^2)
\label{ABCD0}
\end{eqnarray}
It follows from Eq.\ (\ref{ABCD0})
\begin{equation}
r_{\uparrow\uparrow}=-\frac{\cos(2\theta)+i[\mbox{sh}(2\gamma_{0}d)-\sin(2\theta)
\mbox{ch}(2\gamma_{0}d)]}{\mbox{ch}(2\gamma_{0}d)-\sin(2\theta)
\mbox{sh}(2\gamma_{0}d)}+{\cal O}(\epsilon^2)\ ,
\label{Rupup}
\end{equation}
and
\begin{equation}
r_{\downarrow\uparrow}=\frac{\epsilon}{2}\frac{\sin(2\theta)[1-\cos(2\theta)]}
{\mbox{ch}(2\gamma_{0}d)-\sin(2\theta)\mbox{sh}(2\gamma_{0}d)}
+{\cal O}(\epsilon^2)\ .
\label{Rdnup}
\end{equation}

\subsection{An electron incident from the spin-down channel}

The reflection amplitudes for an electron incident from the spin-down
channel can be solved similarly. Now the wavefunction in the electrode
is given by
\begin{equation}
\Psi_{E}(x)=\frac{1}{\sqrt{2}}\vert\downarrow\rangle\otimes\left(
\begin{array}{c}
1\\
1
\end{array}
\right)+
\frac{r_{\downarrow\downarrow}}{\sqrt{2}}\vert\downarrow\rangle\otimes\left(
\begin{array}{c}
1\\
-1
\end{array}
\right)
+
\frac{r_{\uparrow\downarrow}}{\sqrt{2}}\vert\uparrow\rangle\otimes\left(
\begin{array}{c}
1\\
1
\end{array}
\right)\ .
\label{Waveelectrode1}
\end{equation}
The forms of the wavefunctions in the pump and barrier remain to be the same.
By some algebra, we arrive at
\begin{equation}
\Psi_{B}(0^{+})
=\frac{1}{\sqrt{2}}
\left(
\begin{array}{c}
r_{\uparrow\downarrow}\Gamma_{-}(2\gamma_{0}d)\\
r_{\uparrow\downarrow}\Gamma_{+}(2\gamma_{0}d)\\
\Gamma_{+}(2\gamma_{0}d)+r_{\downarrow\downarrow}\Gamma_{-}(2\gamma_{0}d)\\
\Gamma_{-}(2\gamma_{0}d)-r_{\downarrow\downarrow}\Gamma_{+}(2\gamma_{0}d)
\end{array}
\right)\ .
\label{PsiP1}
\end{equation}
Equating Eq.\ (\ref{PsiM}) with Eq.\ (\ref{PsiP1}), we obtain
\begin{eqnarray}
&&\left(\begin{array}{cc}
-i\cos^2\theta&\sin\theta\cos\theta\\
-\sin\theta\cos\theta&-i\sin^2\theta
\end{array}\right)
\left(
\begin{array}{c}
\Gamma_{-}\\
-\Gamma_{+}
\end{array}
\right)
\frac{r_{\uparrow\downarrow}}
{i\frac{R_{0}}{v_{\mbox{\tiny F}}}\cos^3\theta\sin\theta}\nonumber\\
&=&\left(
\begin{array}{c}
\Gamma_{+}(2\gamma_{0}d)+r_{\downarrow\downarrow}\Gamma_{-}(2\gamma_{0}d)\\
\Gamma_{-}(2\gamma_{0}d)-r_{\downarrow\downarrow}\Gamma_{+}(2\gamma_{0}d)
\end{array}
\right)+{\cal O}(\epsilon^2)
\label{ABCD1}
\end{eqnarray}
It follows from Eq.\ (\ref{ABCD1})
\begin{equation}
r_{\downarrow\downarrow}=\frac{\cos(2\theta)-i[\mbox{sh}(2\gamma_{0}d)-\sin(2\theta)
\mbox{ch}(2\gamma_{0}d)]}{\mbox{ch}(2\gamma_{0}d)-\sin(2\theta)
\mbox{sh}(2\gamma_{0}d)}+{\cal O}(\epsilon^2)\ ,
\label{Rdndn}
\end{equation}
and
\begin{equation}
r_{\uparrow\downarrow}=-\frac{1}{2}\frac{\sin(2\theta)[1+\cos(2\theta)]}
{\mbox{ch}(2\gamma_{0}d)-\sin(2\theta)\mbox{sh}(2\gamma_{0}d)}
\left(\frac{R_{0}}{v_{\mbox{\tiny F}}}\right)+{\cal O}(\epsilon^2)\ .
\label{Rupdn}
\end{equation}

\subsection{A verification of the result}

The total Hamiltonian of the system is invariant under the transformation
\begin{equation}
(-i\hat{s}_{y}\hat{\sigma}_{z})H(-\tilde{k}_y)(i\hat{s}_{y}\hat{\sigma}_{z})
=H(\tilde{k}_y)\ ,
\end{equation}
so the corresponding transformation of Eq.\ (\ref{Waveelectrode})
\begin{eqnarray}
(-i\hat{s}_{y}\hat{\sigma}_{z})&&\Psi_{E}(x)\vert_{\tilde{k}_{y}\rightarrow -\tilde{k}_{y}}
=\frac{1}{\sqrt{2}}\vert\downarrow\rangle\otimes\left(
\begin{array}{c}
1\\
1
\end{array}
\right)\nonumber\\
&&+\frac{r_{\uparrow\uparrow}\vert_{\tilde{k}_{y}\rightarrow -\tilde{k}_{y}}}{\sqrt{2}}\vert\downarrow\rangle\otimes\left(
\begin{array}{c}
1\\
-1
\end{array}
\right)\nonumber\\
&&-\frac{r_{\downarrow\uparrow}\vert_{\tilde{k}_{y}\rightarrow -\tilde{k}_{y}}}{\sqrt{2}}\vert\uparrow\rangle\otimes\left(
\begin{array}{c}
1\\
1
\end{array}
\right)\ ,
\end{eqnarray}
must also be an eigenstate of $H(\tilde{k}_{y})$.
This result tells that an electron incident from the spin-down channel
will be reflected into the spin-down channel with amplitude $r_{\uparrow\uparrow}
\vert_{\tilde{k}_{y}\rightarrow -\tilde{k}_{y}}$, and also into the spin-up channel
with amplitude $-r_{\downarrow\uparrow}\vert_{\tilde{k}_{y}\rightarrow -\tilde{k}_{y}}$.
Comparing it with Eq.\ (\ref{Waveelectrode1})
and noticing that $\tilde{k}_{y}\rightarrow -\tilde{k}_{y}$ is equivalent to
$2\theta\rightarrow(\pi-2\theta)$, we find immediately the following relations
\begin{equation}
r_{\downarrow\downarrow}=r_{\uparrow\uparrow}\vert_{2\theta\rightarrow(\pi-2\theta)}\ ,
\end{equation}
and
\begin{equation}
r_{\uparrow\downarrow}=-r_{\downarrow\uparrow}\vert_{2\theta\rightarrow(\pi-2\theta)}\ .
\end{equation}
The reflection amplitudes given in Eqs.\ (\ref{Rupup}), (\ref{Rdnup}),
(\ref{Rdndn}), and (\ref{Rupdn}) apparently satisfy these relations.

%\appendix
\section{CALCULATION OF THE REFLECTION AMPLITUDES FOR A HALF-METALLIC ELECTRODE}

To simulate the half-metallic electrode, the Hamiltonian of the electrode is
taken to be $H_{E}=v'_{\mbox{\tiny F}}(k_{y})k_x\hat{s}_{z}\hat{\sigma}_{x}+
V_{1}(\hat{1}\mp\hat{s}_{z})\hat{\sigma}_{z}/2$. Consider first the case, where
the spin polarization of the electrode is parallel to the $z$-axis.
Now the Hamiltonian of the potential barrier becomes
\begin{equation}
H_{B}=H_{E}+\left(
\begin{array}{cc}
0&0\\
0&V_{1}\hat{\sigma}_{z}
\end{array}
\right)\ .
\end{equation}
For simplicity,
we will take $V_{1}\rightarrow\infty$ limit in the final result.
In this case, an electron incident from the spin-up channel
has some probability to be reflected into the spin-down channel
from the pump, but the reflected wave will decay quickly
to $0$ within the barrier, and has no chance
to reach the electrode. Therefore, $r_{\downarrow\uparrow}\equiv 0$
and the wavefunction in the electrode becomes
\begin{equation}
\Psi_{E}(x)
=\frac{1}{\sqrt{2}}\vert\uparrow\rangle\otimes\left(
\begin{array}{c}
1\\
-1
\end{array}
\right)+\frac{r_{\uparrow\uparrow}}{\sqrt{2}}\vert\uparrow\rangle\otimes\left(
\begin{array}{c}
1\\
1
\end{array}
\right)\ .
\label{WaveelectrodeBlock}
\end{equation}
The wavefunction in the potential barrier can be written as
\begin{widetext}
\begin{eqnarray}
\Psi_{B}(x)&=&\frac{C_{1}}{\sqrt{2}}\vert\uparrow\rangle\otimes\left(
\begin{array}{c}
1\\
-i
\end{array}
\right)
e^{\gamma_{0}x}+\frac{C_{2}}{\sqrt{2}}\vert\uparrow\rangle\otimes\left(
\begin{array}{c}
1\\
i
\end{array}
\right)
e^{-\gamma_{0}x}
+
\frac{C_{3}}{\sqrt{2}}\vert\downarrow\rangle\otimes\left(
\begin{array}{c}
1\\
-i
\end{array}
\right)
e^{-\gamma_{1}x}\ ,
\label{WaveBarrierBlock}
\end{eqnarray}
\end{widetext}
where $\gamma_{0}=V_{0}/\hbar v'_{\mbox{\tiny F}}$
and $\gamma_{1}=V_{1}/\hbar v'_{\mbox{\tiny F}}
\rightarrow\infty$. The wavefunction in the pump remains to be same.
Repeating the same calculation as in sec.\ II, it is straightforward
to obtain
\begin{equation}
r_{\uparrow\uparrow}=-\frac{\cos(2\theta)+i[\mbox{sh}(2\gamma_{0}d)-\sin(2\theta)
\mbox{ch}(2\gamma_{0}d)]}{\mbox{ch}(2\gamma_{0}d)-\sin(2\theta)
\mbox{sh}(2\gamma_{0}d)}+{\cal O}(\epsilon^2)\ .
\label{Rupup_down}
\end{equation}
This expression is identical in form to Eq.\ (\ref{Rupup})
obtained in Sec.\ II for a nonmagnetic electrode.
However, an important difference is $r_{\downarrow\uparrow}\equiv 0$. As a result,
$\vert r_{\uparrow\uparrow}\vert^2\equiv(1-\vert r_{\downarrow\uparrow}\vert^2)\equiv 1$
up to any order in $\epsilon$. This means that ${\cal O}(\epsilon^2)$
in Eq.\ (\ref{Rupup_down})
must be a correction only to the argument of $r_{\uparrow\uparrow}$, which
as discussed in the manuscript,
will not modify the orbit of $r_{\uparrow\uparrow}$ on the complex plane.
The orbit is always a unit circle for $\vert k_{y}\vert <k_{y}^{c}$,
and the amount of charge pumped per cycle by the $k_y$ state
is quantized to $\Delta q(k_{y})=-eC_{+}$. Similarly, for
the spin polarization of the electrode antiparallel to the $z$-axis,
the charge pumped per cycle is $\Delta q(k_{y})=-eC_{-}$.

%\end{CJK}

\end{document}